\documentclass[twocolumn,prd]{revtex4}
\usepackage{graphicx}
\usepackage{color}
\usepackage{balance}
\usepackage[skip=3pt]{caption}
\newcommand{\One}{1\:\!\!\!\!1}
\renewcommand{\ap}{\'{a}}
\newcommand{\ag}{\`{a}}
\newcommand{\ep}{\'{e}}
\newcommand{\ip}{\'{\i}}
\newcommand{\op}{\'{o}}
\newcommand{\up}{\'{u}}
\newcommand{\at}{\~{a}}
\newcommand{\ot}{\~{o}}
\newcommand{\ah}{\^{a}}
\newcommand{\eh}{\^{e}}
\newcommand{\cd}{\c{c}}
\renewcommand{\ao}{{\at}o}
\newcommand{\ote}{{\ot}e}
\newcommand{\oes}{{\ot}es}
\newcommand{\cao}{\cd{\at}o}
\newcommand{\coes}{\cd{\ot}es}
\newcommand{\hsp}[1]{\hspace*{#1mm}}
\newcommand{\rbt}[1]{\raisebox{1ex}{\underline{\tiny #1}}}

\begin{document}
\font\myfont=cmbx12 at 15pt
\title{\mbox{\myfont
\hsp{-11}Espectroscopia Mes{\op}nica Moderna: 
o Papel Fundamental da Unitariedade}}
\author{Eef van Beveren$^{a}$ e George Rupp$^{b}$} 
\affiliation{
$^a$Centro de F{\ip}sica da UC, Departamento de F{\ip}sica,
Universidade de Coimbra; eef@uc.pt \\
\mbox{${^b}$Centro de F{\ip}sica e Engenharia de Materiais Avan{\cd}ados,
Instituto Superior T{\ep}cnico; george@tecnico.ulisboa.pt}
}
\thispagestyle{empty}
\begin{abstract}
The importance of implementing unitarity constraints in meson spectroscopy
is very briefly outlined for Portuguese students of engineering sciences
and therefore non-experts in the field. After explaining the profound
differences between meson spectroscopy and atomic spectroscopy, attention
is paid to the shortcomings of standard Breit-Wigner parametrisations in
the case of broad and/or overlapping resonances.
Finally, the manifestly unitary Resonance-Spectrum-Expansion model, which
lies at the heart of a recent invited review paper by the present
authors, is graphically presented, together with a simple yet typical
application to the long-controversial $K_0^\star(700)$ resonance.
\end{abstract}
\maketitle
\section{Introdu\cao}
A Espectroscopia Mes{\op}nica tem como objectivo principal descrever de forma
sistem{\ap}tica as massas dos estados fundamentais e excitados de mes\oes\
observados em experi{\eh}ncias nos v{\ap}rios aceleradores de part{\ip}culas
por esse mundo fora. 
Os mes\oes\ s\ao\ sistemas de spin inteiro que consistem em dois fermi\oes,
um quark ($q$) e um antiquark ($\bar{q}$).
Quando a massa e os n\'{u}meros qu\^{a}nticos o permitem, um mes\ao\
desintegra-se rapidamente em pares de mes\oes\ mais leves por meio da cria\cao\
de um par $q\bar{q}$, tratando-se de um estado energeticamente favorecido.
Neste caso, falamos de uma \em resson\^{a}ncia mes\'{o}nica, \em
com largura inversamente proporcional ao seu tempo de vida.
A mes\oes\ que n\ao\ se podem desintegrar desta forma chamamos, ignorando
decaimentos muito mais lentos, \em estados ligados. \em

\begin{figure}[!t]
\begin{center}
\includegraphics[trim = 0mm 10mm 0mm 0mm,clip,width=8cm,angle=0]
{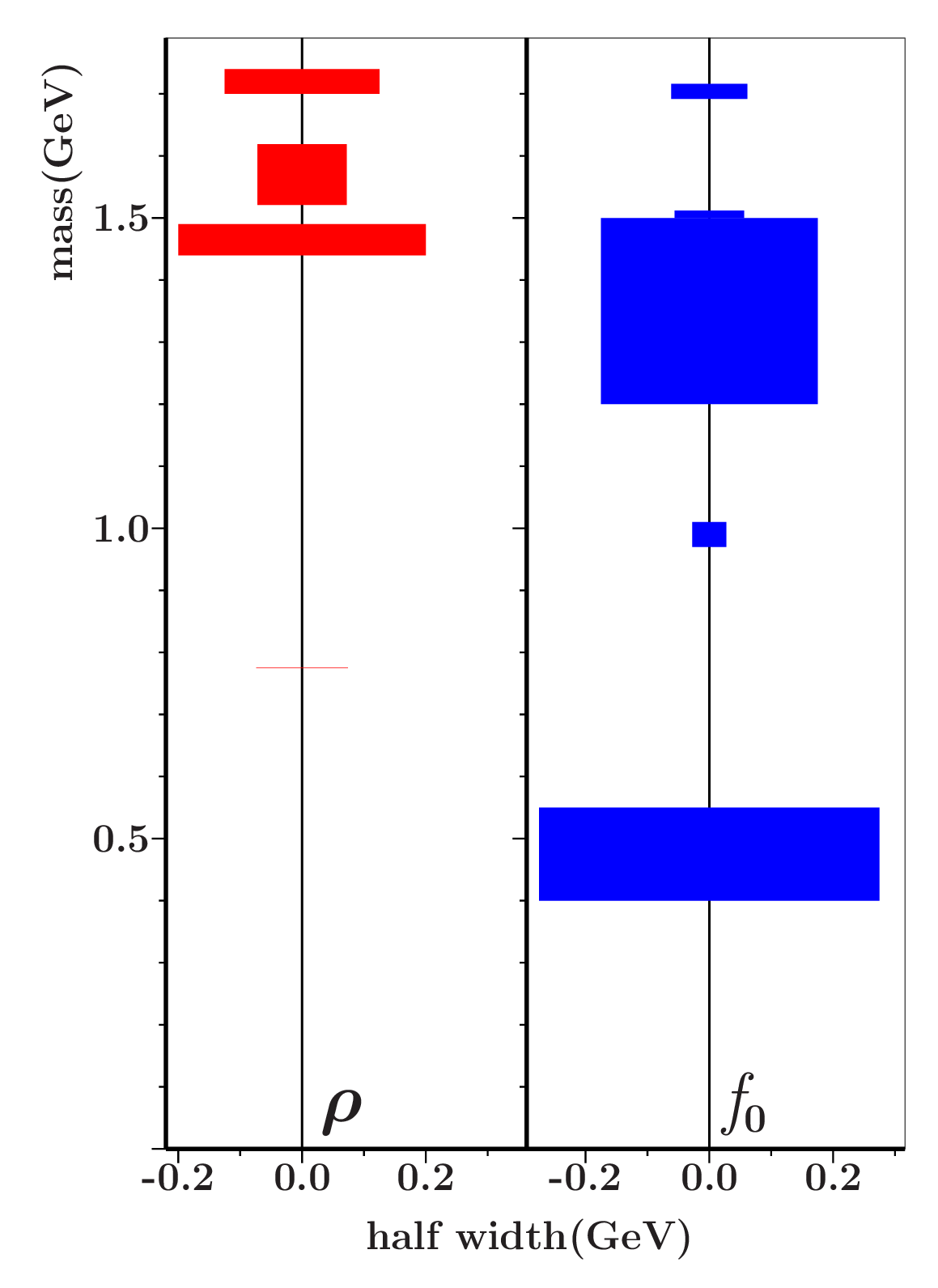}
\end{center}
\caption{$\rho(770)$, $f_0(500)$ e as suas excita\coes.}
\label{rhof0}
\end{figure}

\begin{figure}[!t]
\begin{center}
\includegraphics[trim = 0mm 0mm 0mm 0mm,clip,width=5.1cm,angle=-90]
{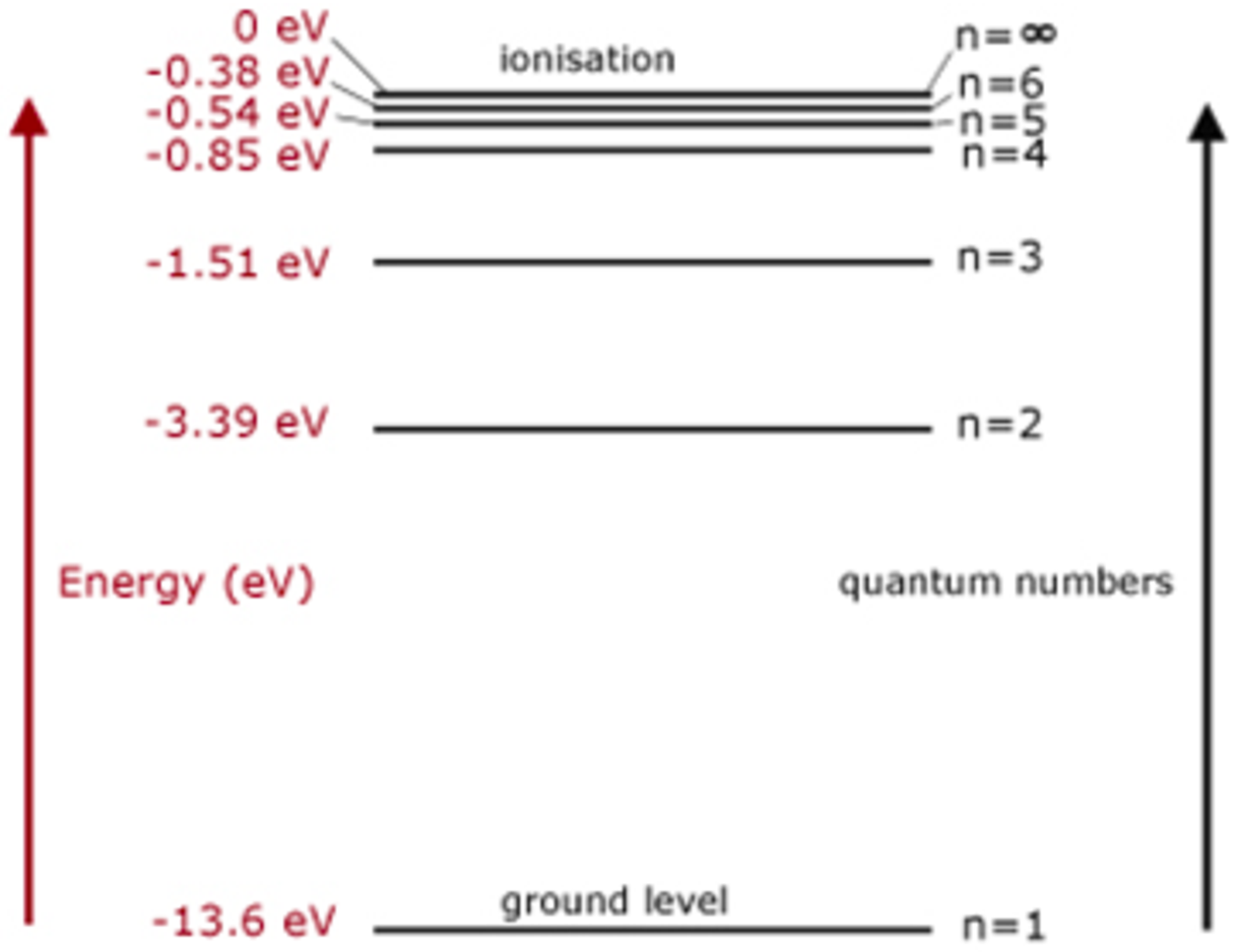}
\end{center}
\caption{\mbox{N{\ip}veis de energia de hidrog{\ep}nio at{\op}mico.}}
\label{hydrogen}
\end{figure}

A for{\cd}a que liga o
par $q\bar{q}$ n\ao\ \ep\ inteiramente conhecida, pois resulta da
Cromodin{\ah}mica Qu{\ah}ntica (QCD), que \ep\ uma teoria n\ao\ sol{\up}vel a
baixas ener\-gias. Mesmo assim, sabemos empiricamente que o respectivo
potencial \ep\ sempre crescente \ag\ medida que a dist{\ah}ncia entre o par
$q\bar{q}$ aumenta, j\ap\ que (anti)quarks nunca foram observados isoladamente.
Por isso se fala num potencial \em confinante, \em caracter{\ip}stica essa que
j\ap\ foi confirmada em simu\-la\coes\ num{\ep}ricas num espa{\cd}o-tempo
discretizado e finito \em c{\ap}lculos na rede, ``LQCD''). \em
Para perceber melhor o potencial confinante conv{\ep}m analisar os espectros
de mes\oes\ com n{\up}meros qu{\ah}nticos diferentes, e tamb{\ep}m
compostos de (anti)quarks dos diversos \em sabores \em ($u$, $d$, $s$, $c$ e
$b$).  No entanto, uma compara\cao\ com a Espectroscopia
At{\op}mica revela logo v{\ap}rias complica\coes, que iremos
ilustrar atrav{\ep}s das Figuras~\ref{rhof0} e \ref{hydrogen}.

Figura~\ref{rhof0} exibe as massas dos mes\oes\ vectorial (spin 1) $\rho(770)$
e escalar (spin 0) $f_0(500)$ com as suas excita\coes\ radiais at\ep\ cerca de
1.7~GeV (1700~MeV). Representamos todos estes mes\oes\ aqui por um
rect{\ah}ngulo, a fim de mostrar a dificuldade em atribuir uma massa precisa
a cada um. A altura dos rect{\ah}ngulos corresponde \ag\ res\-pectiva incerteza
experimental na massa, na escala da ordenada e segundo as tabelas do Particle
Data Group (PDG), actualizadas anual e publicadas \cite{PDG2020} bienalmente.
As correspondentes dimens\oes\ na abscissa d\ao\ as larguras de decaimento dos
v{\ap}rios mes\oes, outra vez na mesma escala que a da ordenada. Como j\ap\
referimos, resson\^{a}ncias mes\'{o}nicas desintegram-se rapidamente e os
estados $\rho$ e $f_0$, que se desintegram ambos em dois pi\oes, ``vivem''
durante t\ao\ pouco tempo ($10^{-24}$ a $10^{-23}$~seg.) que at\ep\ se torna
discut{\ip}vel atribuir-lhes uma massa. Torna-se assim indispens{\ap}vel
definir de uma forma n\ao\ amb{\ip}gua o que se entende por massa no caso de
part{\ip}culas altamente inst{\ap}veis. Esta proble\-m{\ap}tica \ep\ o ponto
central do presente artigo e ser\ap\ abordado nas sec\coes\ seguintes.

Falta ainda chamar a aten\cao\ para mais
diferen{\cd}as entre os, \ag\ primeira vista, incompreens{\ip}veis espectros
dos $\rho$ e $f_0$, quando comparados com o simples espectro de hidrog{\ep}nio
at{\op}mico na Figura~\ref{hydrogen} (da Ref.\rbt{a}~\cite{brilliant}). Neste
caso, a incerteza nos n{\ip}veis de energia \ep\ quase nula, pois tanto
te{\op}rica como experimentalmente se trata de um sistema extremamente bem
conhecido e medido. Tamb{\ep}m a largura de decaimento dos n{\ip}veis \ep\
totalmente desprez{\ap}vel em compara\cao\ com as separa\coes\ entre eles, o
que ma\-ni\-festamente n\ao\ se aplica aos referidos mes\oes. Finalmente, os
n{\ip}veis de energia de hidrog{\ep}nio t{\eh}m uma refer{\eh}ncia relativa na
energia de desintegra\cao\ num prot\ao\ e um electr\ao, enquanto os mes\oes\
n\ao\ se podem desintegrar num quark e um antiquark, pelo que \ep\
necess{\ap}rio usar a massa total do mes\ao\ como refer{\eh}ncia.
\section{\mbox{Resson{\ah}ncias Breit-Wigner}}
\begin{figure}[!b]
\begin{center}
\includegraphics[trim = 0mm 0mm 0mm 0mm,clip,width=5.1cm,angle=-90]
{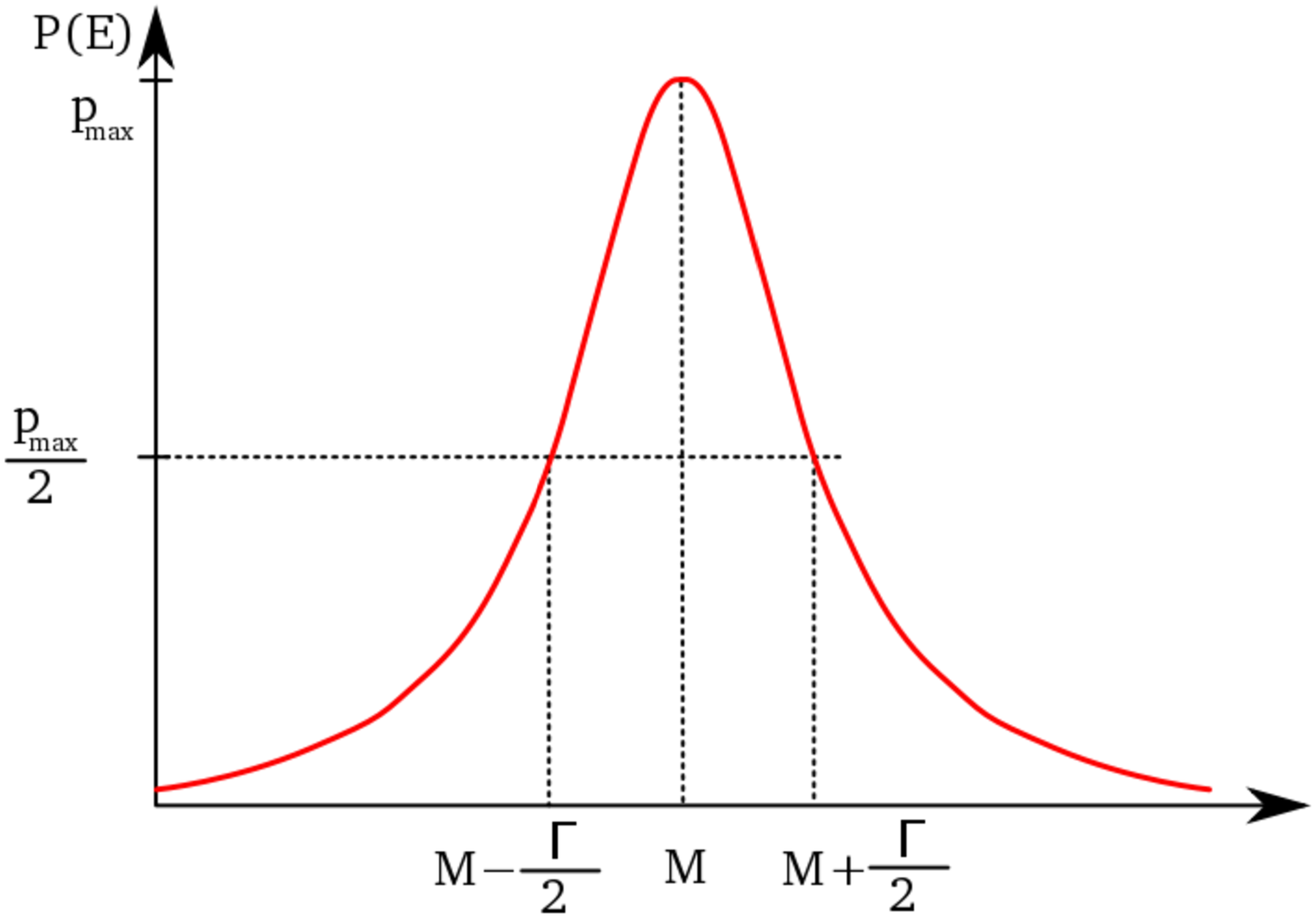}
\end{center}
\caption{Resson{\ah}ncia Breit-Wigner \cite{BW}.}
\label{bw}
\end{figure}

A forma tradicional de modelar uma resson{\ah}ncia isolada
\ep\ atrav{\ep}s de
uma parametriza\cao\ \em Breit-Wigner \em \/(BW), que mostramos
esquematicamente na Figura~\ref{bw} (da Ref.\rbt{a}~\cite{BW}). A
respectiva express\ao\ como fun\cao\ da energia \ep\  \\[-5mm]
\begin{equation}
P(E) \propto \left[(E-M)^2 + \Gamma^2/4\right]^{-1} \; , \\[-2mm]
\label{bw-par}
\end{equation}
onde $M$ \ep\ a massa e $\Gamma$ a largura da resson{\ah}ncia. Verifica-se
facilmente que a curva na Figura~\ref{bw} atinge o seu m{\ap}ximo 
$P_{\mbox{\scriptsize max}}$ quando $E\!=\!M$ e $\Gamma$ \ep\ mesmo a
sua largura a meio de $P_{\mbox{\scriptsize max}}$. Se permitirmos
energias complexas, $P(E)$ vai para infinito se $E\to M\pm i\Gamma/2$, ao
que se chama \em p{\op}los \em \/nesta fun\cao. Isto pode parecer apenas uma
curiosidade acad{\ep}mica, mas n\ao\ o \ep, como veremos na sec\cao\ seguinte
sobre \em unitariedade. \em De qualquer modo, a parametriza\cao\ BW de uma
resson{\ah}ncia s\op\ \ep\ boa quando esta estiver bastante afastada tanto de
poss{\ip}veis outras resson{\ah}ncias como de um limiar de decaimento, que
corresponde \ag\ soma das massas das part{\ip}culas em que a resson{\ah}ncia
se desintegra. Por exemplo, o $\rho(770)$ tem, como {\up}nico limiar, a energia
de 280~MeV, ou seja, a soma das massas de dois pi\oes\ ($\pi\pi$). Uma vez que
o intervalo de energia entre o $\rho(770)$ e esse limiar \ep\ mais de 3 vezes a
largura total do $\rho(770)$, enquanto a dist{\ah}ncia at\ep\ \ag\ primeira
excita\cao\ $\rho(1450)$ (segundo o PDG \cite{PDG2020}, Figura~\ref{rhof0})
\ep\ ainda maior, a parametriza\cao\ BW afigura-se razo{\ap}vel neste caso.
Por{\ep}m, isso j\ap\ n\ao\ se pode dizer da resson{\ah}ncia escalar
$f_0(500)$ (Ref.\rbt{a}~\cite{PDG2020}, Figura~\ref{rhof0}), que fica a menos
de metade da sua massa e largura m{\ep}dia 
do mesmo limiar ($\pi\pi$). Para
este e tamb{\ep}m outros mes\oes\ na Figura~\ref{rhof0} imp\ote-se
evidentemente uma descri\cao\ mais realista.

\section{\mbox{Modelo RSE e Unitariedade}}
H\ap\ muito anos, conseguimos descrever \cite{beveren86} a
resson{\ah}ncia $f_0(500)$ e os outros mes\oes\ escalares leves $f_0(980)$,
$K_0^\star(700)$ e $a_0(980)$ \cite{PDG2020} num modelo de quarks
\cite{beveren80,beveren83} desenvolvido com colegas na Universidade de
Nijmegen, Pa{\ip}ses Baixos. A ess{\eh}ncia do modelo era a descri\cao, com uma
s\'{o} express\ao\ incluindo todos os poss\'{\i}veis estados ligados e
resson\^{a}ncias mes{\op}nicos com os mesmos n\'{u}meros qu\^{a}nticos e
conte\'{u}do de sabor, atrav{\ep}s de um tratamento em p\ep\ de igualdade do
confinamento de quarks e o decaimento forte. Este formalismo \ep\ muito
t{\ep}cnico e fica fora do {\ah}mbito do presente breve resumo.
No entanto, conv{\ep}m ainda mencionar que o mesmo modelo descrevia com sucesso
os ent\ao\ conhecidos espectros de charm{\op}nio (estados $c\bar{c}$) e
bottom{\op}nio ($b\bar{b}$) \cite{beveren80}, al{\ep}m de prever
\cite{beveren83}, entre muitos outros mes\oes, uma resson{\ah}ncia $\rho(1250)$
(ver tamb{\ep}m a Ref.\rbt{a}~\cite{hammoud20})
no largo intervalo de energia onde o PDG apenas identifica um $\rho(1450)$
(ver Figura~\ref{rhof0}). 
Outra grande surpresa na Ref.\rbt{a}~\cite{beveren86} foi
a descri\cao\ dos mes\oes\ escalares leves como resson{\ah}ncias din{\ah}micas
e n\ao\ simplesmente ligadas ao espectro do potencial confinante, nem como
estados do tipo $qq\bar{q}\bar{q}$ \em (``tetraquarks''). \em 

\begin{figure}[!t]
\begin{tabular}{cccc}
\hsp{-3}
\includegraphics[trim = 15mm 0mm 0mm 0mm,clip,height=1.35cm,angle=0]
{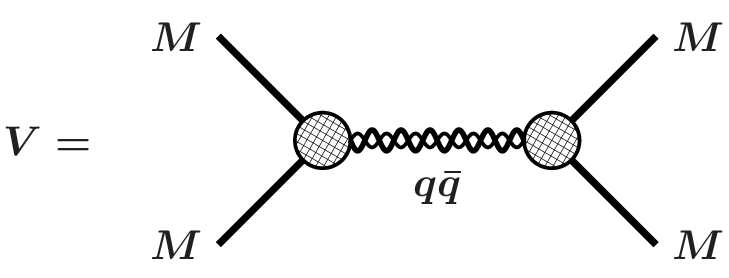} &
\raisebox{5mm}{$\!\!+\!\!$} &
\includegraphics[trim = 24mm 0mm 0mm 0mm,clip,height=1.35cm,angle=0]
{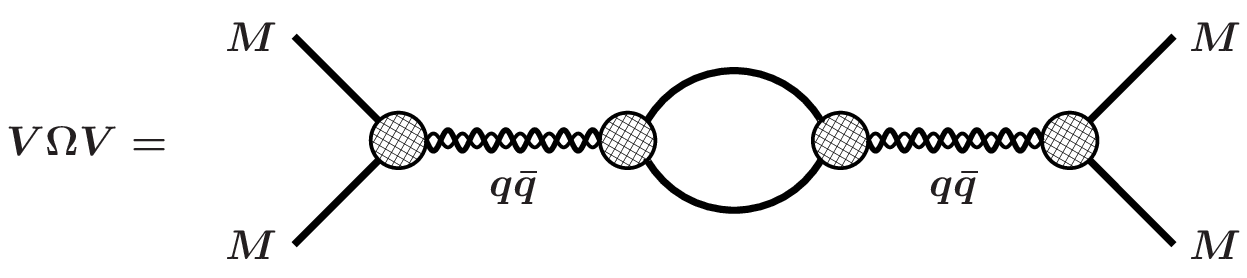} &
\raisebox{5mm}{$\!\!\!+\;\ldots$} \\[2mm]
\end{tabular}
\caption{Representa\cao\ esquem{\ap}tica do modelo RSE.}
\label{rse}
\end{figure}

Muitos anos depois, um conhecido f{\ip}sico experimental desafiou-nos para 
desenvolver um modelo de quarks semelhante, mas de mais f{\ap}cil utiliza\cao,
o que nos levou ao modelo {\bf RSE} \em (``Resonance Spectrum Expansion'') \em
\cite{beveren06}. A ideia b{\ap}sica \ep\ que resson{\ah}ncias
mes{\op}nicas podem ser consideradas estados interm{\ep}dios em colis\oes\
entre dois mes\oes\ mais leves, mas que ao mesmo tempo t{\eh}m a identidade de
estados $q\bar{q}$ com os mesmos n{\up}meros qu{\ah}nticos (spin, \em paridade,
\em \ldots). Isso \ep\ simbolicamente retratado na Figura~\ref{rse}, onde o
primeiro diagrama representa o processo mais simples, com os dois mes\oes\ se
``fundindo'' no primeiro v{\ep}rtice, propagando-se depois como um conjunto de
estados $q\bar{q}$ e desintegrando-se novamente em dois mes\oes\ no segundo
v{\ep}rtice. Os processos nos v{\ep}rtices s\ao\ a aniquila\cao\ e cria\cao\
de um par $q\bar{q}$, respectivamente. O segundo diagrama na Figura~\ref{rse}
corresponde a um processo qu{\ah}ntico de ordem superior, envolvendo duas
aniquila\coes\ e duas cria\coes\ de $q\bar{q}$. Os pontinhos representam
\em todas \em \/as ordens superiores, que pela estrutura dos diagramas
permitem somar tudo, alg{\ep}brica e analiticamente, para se obter uma
express\ao\ fechada para o que se chama a \em amplitude \em \/deste processo.
Isto equivale a resolver uma equa\cao\ do tipo \em Schr\"{o}dinger. \em Com a
tal amplitude constr{\op}i-se directamente a chamada \em matriz \em
{\boldmath$S$}, que \ep\ a grandeza qu{\ah}ntica mais geral para descrever
processos de colis\oes\ entre part{\ip}culas.

Observe que, embora todas as linhas externas dos diagramas na Figura~\ref{rse}
sejam rotuladas com a mesma letra $M$, os pares de mes\oes\ dos lados esquerdo
e direito n\ao\ precisam de ser iguais. Por exemplo, na dispers\ao\ de pi\oes\
pode-se formar uma resson\^{a}ncia da fam\'{\i}lia dos $\rho$ pela
aniquila\cao\ de um par $q\bar{q}$ leve. Caso a massa total o permita, esta
resson\^{a}ncia pode tamb{\ep}m desintegrar-se num par de ka\oes\ (KK)
atrav\'{e}s da cria\cao\ de um par $s\bar{s}$. A matriz {\boldmath$S$} descreve
assim n\ao\ apenas a dispers\ao\ de dois pi\oes, mas tamb\'{e}m a sua
transforma\cao\ em pares de outros mes\oes\ e ainda a dispers\ao\ directa
destes mes\oes.

A enorme vantagem de se ter uma forma expl{\ip}cita da matriz {\boldmath$S$}
\ep\ que a mesma cont{\ep}m \em toda \em \/a informa\cao\ do processo,
incluindo limiares de decaimento, estados ligados,
resson{\ah}ncias simples e resson{\ah}ncias din{\ah}micas. Estas duas
{\up}ltimas correspondem a singularidades complexas em {\boldmath$S$}, os tais
\em p{\op}los \em \/acima referidos, mas agora bem realistas e sem
aproxima\coes\ do tipo BW para cada uma destas resson{\ah}ncias separadamente.
A matriz {\boldmath$S$} assim derivada \ep\ manifestamente sim{\ep}trica e
unit{\ap}ria, propriedades matem{\ap}ticas essas que garantem invari{\ah}ncia
de revers\ao\ de tempo e conserva\cao\ de probabilidade, respectivamente. A
unitariedade de {\boldmath$S$} implica
$\left(\mbox{\boldmath$S$}^\star\right)^{\!T}\!\mbox{\boldmath$S$}=\One$,
onde o asterisco significa complexo-conjugado e  o''$T$'' transposi\cao.
\begin{figure}[!b]
\begin{center}
\includegraphics[trim = 61mm 184mm 49mm 18mm,clip,width=8cm,angle=0]
{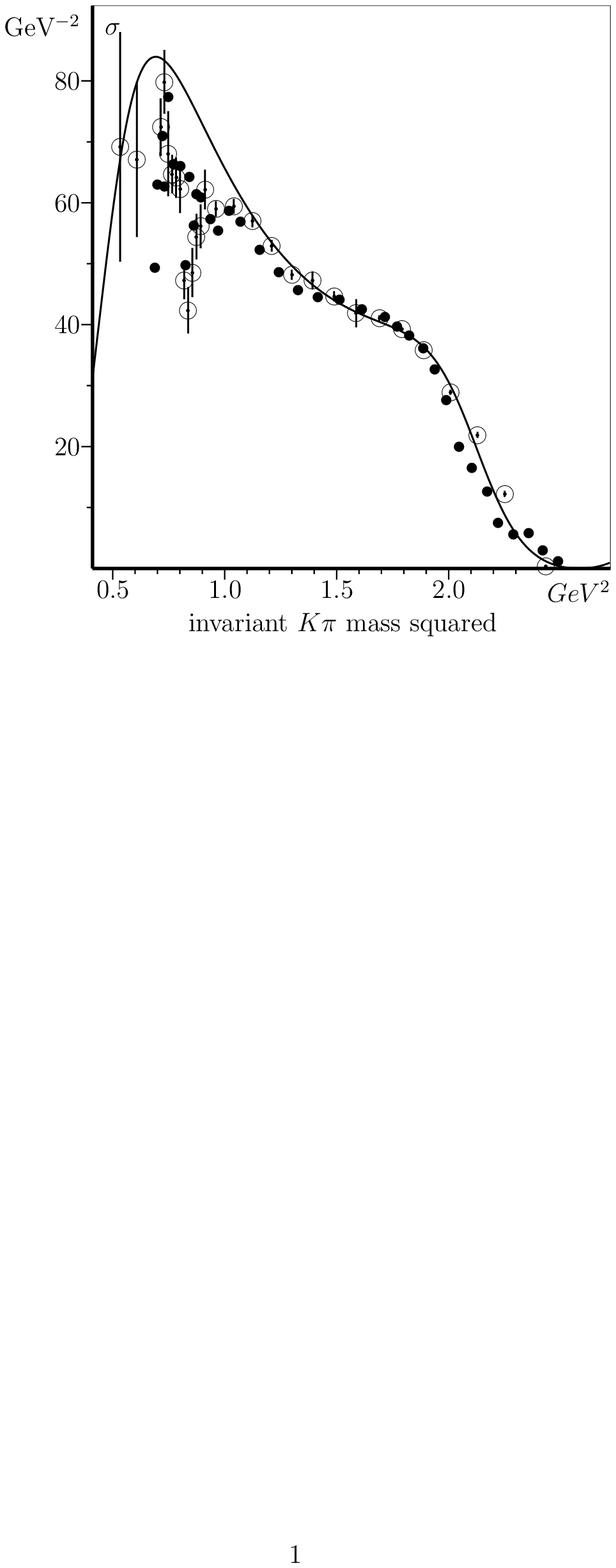} 
\end{center}
\caption{Sec\cao\ eficaz $K\pi$ em onda $S$ \cite{beveren01}.}
\label{kappa}
\end{figure}

Uma aplica\cao\ muito simples \cite{beveren01} do modelo RSE, com o {\up}nico
modo de decaimento $K\pi$ em onda $S$ ($l\!=\!0$), resulta numa not{\ap}vel
descri\cao\ da respectiva sec\cao\ eficaz experimental at\ep\ $E\!=\!1.6$~GeV
(ver Figura~\ref{kappa}), com muito poucos par{\ah}metros. Al{\ep}m disso,
consegue identificar um p{\op}lo din{\ah}mico de energia complexa a
$(714-i228)$~MeV, correspondente \ag\ resson{\ah}ncia escalar $K_0^\star(700)$
\cite{PDG2020}.

Desde ent\ao\ fizemos muitas outras aplica\coes\ do modelo RSE \ag\
espectroscopia mes{\op}nica, descrevendo resson{\ah}ncias mais complicadas,
que envolvem v{\ap}rios modos diferentes de decaimento. Exemplos
bem-sucedidos, al{\ep}m dos mes\oes\ escalares leves e nalguns casos j\ap\
confirmados pela LQCD, s\ao\ os controversos
mes\oes\ $D_{s0}^\star(2317)$ \cite{beveren03}, $D_0^\star(2300)$,
$D_{s1}(2460)$, $D_1(2430)$, $D_{sJ}(2860)$, $\chi_{c1}(3872)$ \cite{coito11},
$\psi(4260)$, $\psi(4660)$ e $\Upsilon(10580)$. Trat{\ap}mos de forma muito
mais detalhada destas e doutras resson{\ah}ncias mes{\op}nicas num recente
artigo de revis\ao\ por convite \cite{beveren20}, que tamb{\ep}m aborda
processos de produ\cao\ e efeitos de limiares, com base no princ{\ip}pio da
unitariedade.

\section{\mbox{Resumo}}
Os resultados das experi\^{e}ncias de colis\oes\ s\ao\ dados
em termos de se\coes\ eficazes ou amplitudes em fun\cao\ da energia total.
Geralmente, a amplitude exibe v\'{a}rias eleva\coes\ que podem indicar
a exist\^{e}ncia de resson\^{a}ncias mes\'{o}nicas nas respectivas energias.
A descri\cao\ desta amplitude por uma s\'{o} express\ao\ para todos os membros
de uma fam\'{\i}lia de mes\oes\ \'{e} claramente mais realista do que uma
parametriza\cao\ um a um do tipo Breit-Wigner. 
\renewcommand{\refname}{Refer{\eh}ncias}

\end{document}